\documentclass{article}

\usepackage{PRIMEarxiv}
\usepackage{float}
\usepackage[utf8]{inputenc} 
\usepackage[T1]{fontenc}    
\usepackage{hyperref}       
\usepackage{url}            
\usepackage{booktabs}       
\usepackage{amsfonts}       
\usepackage{nicefrac}       
\usepackage{microtype}      
\usepackage{lipsum}
\usepackage{fancyhdr}       
\usepackage{graphicx}       
\graphicspath{{media/}}     

\title{Mining comorbidities: a brief survey
}

\author{
  Giovanna Maria Dimitri \\
  Department of Computer Science, University of Cambridge \\
  Cambridge, UK \\
}

\begin{document}
\maketitle

\begin{abstract}
In this manuscript we will present a brief overview of the comorbidity concept. We will start by laying its foundations and its definitions and then describing the role that machine learning can hold in mining and defining it. The purpose of this short survey is to present a brief overview of the definition of comorbidity as a concept, and showing some of the latest applications and potentialities for the application of natural language processing and text mining techniques.
\end{abstract}

\keywords{Text Mining \and Machine Learning \and Comorbidites}

\section{Introduction to Comorbidity and Definition}
\label{chapter-1-comorbidity}

\begin{figure}[htp]
    \centering
    \includegraphics[scale=0.4]{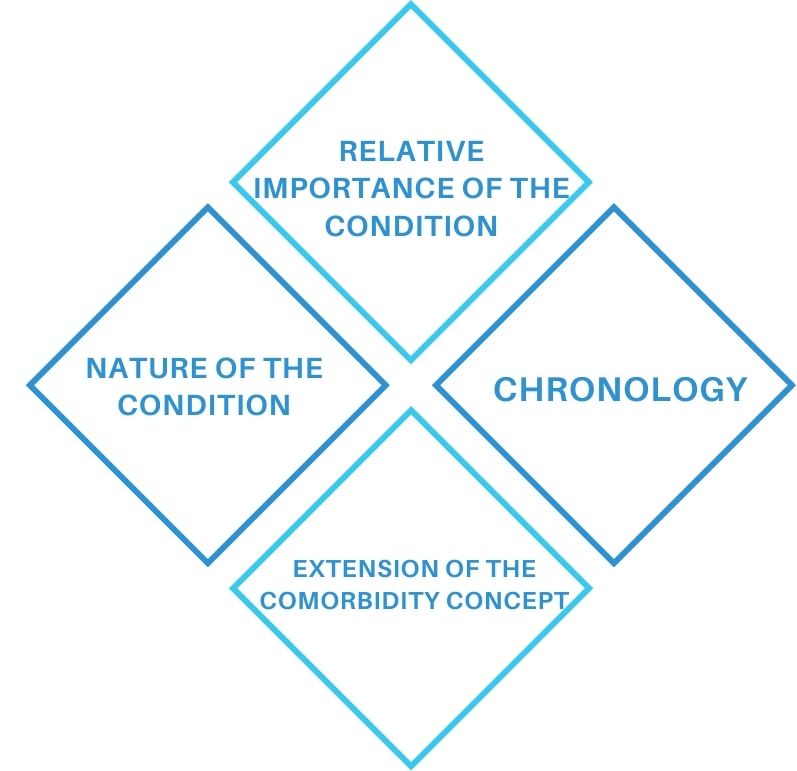}
    \caption{The 4 main dimensions of the concept of Comorbidity}
    \label{fig:Comorbidity}
\end{figure}

The definition of comorbidity has been at the centre of a debate for several years, both in the medical and life sciences literature. To
provide a proper definition of comorbidity we could possibly critically analyse and start from the definition given in
\cite{1}. In general, there exists, in fact, a lack of consensus \cite{1} in the way the concept of
comorbidity is defined and measured. It is
possible to start summarizing how the concept of comorbidity can be
conceptualized in several different ways \cite{2} \cite{3}, however all of them built on a core concept: the presence of more than one distinct condition of morbidity in an individual. In particular we could identify the distinctive characteristics of comorbidity lying in 4 main dimensions summarized also in Figure \ref{fig:Comorbidity} \cite{1}:

\begin{enumerate}
\def\labelenumi{\arabic{enumi}.}
\item
  \textbf{Nature of health condition}: the first important point to take
  into consideration is what the term condition stands for. Indeed, it
  can potentially hold various meanings including disease, disorder, illness or
  health problems. Classifying and defining such different cases is
  crucial to provide a clear conceptualization of the term
  comorbidity. The classification of the condition is therefore the first action to be considered when dealing with the classification of the comorbidity concept. 
\item
  \textbf{The relative importance of the conditions}: this aspect concerns a critical research question arising in the context of comorbidities. When two or more clinical conditions are present, which one
  should be defined as the index (that is the main disease an individual
  holds) and which ones as the co-morbidities associated to the first one?
  The answer to this may vary with the research question or the
  particular disease that we are considering, as well as with the diseases onset chronology which is also described in the following point. This is the reason why,
  together with the concept of co-morbidity, the concept of
  multimorbidity has been developed to describe the \textit{co-occurrence of
  multiple chronic or acute diseases and medical conditions within one
  person} \cite{4}. The case of multimorbidity is in fact an extension of the comorbidity concept where there is no necessity of
  defining an index disease condition with respect to a comorbid
  condition \cite{1}. All of the simultaneous diseases are in fact considered as part of the same multilayered pathological conditions of an individual with no prevalence given to one with respect to another.
\item
  \textbf{Chronology}: not only it is interesting to understand which
  diseases co-appear in a patient, but is also interesting to understand
  the time span and sequence with which two events appear. In fact, even
  if two diseases appear in the same patient at two different points of
  time, it is interesting to study there behaviour from an etiological
  point of view \cite{1}. The temporal dimension of comorbidities should in fact be taken into account when defining the hierarchy of comorbidities and pathological conditions of a patients, and when considering their relative importance.   
\end{enumerate}

For what concerns the quantification of the concept of comorbidity this is not an easy  task. For this purpose many indexes have been
  proposed such as, for example, the Cumulative Illness Rating Scale
  (CIRS) \cite{linn1968cumulative}, the Index of Coexisting Disease (ICED) and the Kaplan Index \cite{sands2006predicting}.
  Depending on the index considered, patients are classified and
  stratified in different categories \cite{1}. A further index that is considered
  is the Patient Complexity. This index takes into account the fact that
  the morbidity burden is not only due to health-related
  characteristics, but also to socioeconomic, cultural and environmental
  features. Capturing and measuring the complexity of socio-economic
  factors involved in comorbidity remains a challenge \cite{1}.
Moreover it is interesting to notice how the different aspects of comorbidity
analysed are all useful to depict an overall view of comorbidity in a
patient. In particular, depending upon who is analysing a patient the
point of view on comorbidity can vary and be susceptible to multiple,
different, definitions when considering the same individual. Not only pathological and physical conditions should in fact be taken into account, but also demographics and several other information on the patients status could actually be extremely important for the definition of the severity of the patients conditions.

The implications of the management and diagnosis of comorbidities can be dramatically appreciated in three main research areas
\cite{1}:

\begin{enumerate}
\def\labelenumi{\arabic{enumi}.}
\item
  \textbf{Clinical care}: when a patient with multiple diseases has to
  be treated, understanding the complexity of his condition, shedding
  lights on his diseases, will help deciding how to treat and cure him.
  In particular building his profile, identifying the index disease,
  will help finding a specialist able to treat and cure him.
\item
  \textbf{Epidemiology and public health}: here the key issue of
  interest with respect to comorbidity is the identification of
  concurrent diseases.
\item
  \textbf{Health service planning and financing}: In this case the main
  issue is to examine how to allocate the resources available to
  treat comorbidities.
\end{enumerate}
\begin{figure}[htp]
    \centering
    \includegraphics[scale=0.3]{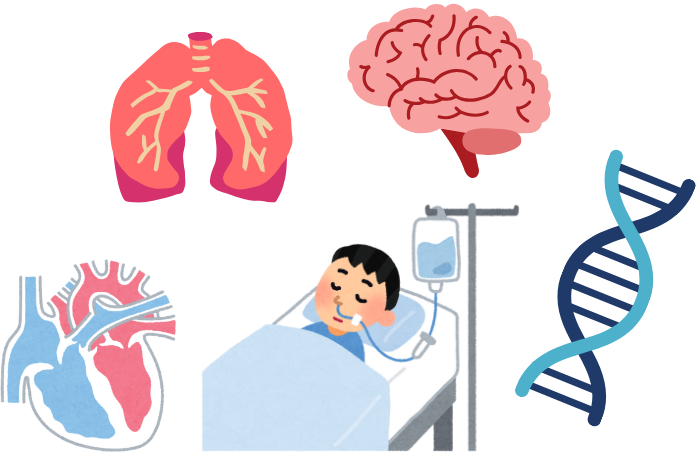}
    \caption{An Infographics showing the presence of comorbidities affecting different organs in a patient}
    \label{fig:ComorbiditiesInfoGraphics}
\end{figure}
Furthermore, it is not easy to understand the epidemiological reasons laying behind the rise of comorbidities in patients. Three main epidemiological factors are mainly identified by the literature \cite{5,1}: chance, selection bias and other types of causal associations.  Even if the first two causes rely mainly on causal links, they still have to
be taken into account when analysing comorbidity since they can lead to
wrong conclusions about causalities. On the other hand regarding causal associations we could further assess the presence of four different type of relationships according to the main cases reported in the literature
\cite{1}:

\begin{enumerate}
\def\labelenumi{\arabic{enumi}.}
\item
  \textbf{Direct causation model}: the presence of one disease is
  directly responsible for the presence of another disease.
\item
  \textbf{Associated risk factor}: the risk factor for one disease is
  correlated with the risk factor of another disease. This means that
  two diseases will more likely occur together if one is present in a
  patient.
\item
  \textbf{Heterogeneity model}: in this case the risk factors are not
  correlated, but each of them can cause either disease.
\item
  \textbf{Independence model}: in this model diagnostic features of
  diseases are simultaneously present and cause a third distinct
  disease.
\end{enumerate}

These four aspects are not mutually exclusive and can be found
simultaneously in a patient.

\section{Mining Comorbidities}
In the following subsections we will collect recent studies on the comorbidities existing in cancer and neurological diseases. We will then go over the description of complex multi-diseases and the use of machine learning for mining them effectively. The reason why we chose to concentrate on these two types of pathologies relies on the need of identifying two major sets of diseases in which comorbidities applies. This brief survey aims at offering, in fact, a short overview on the topic, to be considered as the point of start of discussion for further investigations and research proposals.

\subsection{Cancer-Specific Disease Comorbidities}\label{mining-cancer-specific-disease-comorbidities-from-a-large-observational-health-database-6}
Comorbidities of cancer-related diseases have a long standing story in research.
In \cite{6} the authors systematically mined and analysed
comorbidities associated to cancer. In particular they extracted
cancer-specific comorbidities from the FDA Adverse Event Reporting
System (FAERS), using a data mining approach \cite{6}. In the FAERS
database there are records of 3.354.043 patients (males and females of
all ages), 1.138 cancers of different types and stages, and 8.974
non-cancer health problems. These data can be analysed in terms of
cancer comorbidity patterns, among different patient populations. The initial population of 3.354.043 patients was stratified based on age
and gender. Then through a network-based approach comorbidity patterns
have been extracted from each group of patients. Once comorbidity
patterns among different groups were identified, the authors analysed
the effect of age and gender on them, and results show how the
relationships between cancer and non-cancer diseases depend mostly on
age and gender, except for some exceptions such as depression, anxiety
and metabolic syndromes whose comorbidity relationships with cancers are
quite stable among all patients. The comorbidity mining approach was
applied by the author to colorectal cancer, detecting the association
with metabolic syndrome components, diabetes and osteoporosis.

Cancer comorbidities were mined in the database using the following
three steps:

\begin{enumerate}
\def\labelenumi{\arabic{enumi}.}
\item
  Application of an association rule mining algorithm on the pairs
  patient-disease, mining strong co-occurrence patterns among disease
  combinations.
\item
  Building a co-morbidity network using the resulting patterns.
\item
  Extract comorbidity for cancers. To do that, the authors initiated a
  random walk on the network from a set of interested cancer nodes,
  ranking the non-cancer diseases with the probabilities of being
  reached by the random walk.
\end{enumerate}

This study is particularly relevant to detect comorbidities related to
cancer, showing the interesting role played by age and gender in
discovering comorbidities. However this method concentrates on finding
comorbidities considering cancer only; therefore it could be interesting
to widen the spectrum of interest of diseases. Moreover the paper
investigates the role of age and gender on the diseases, and a further
analysis could focus on the impact of other collateral variables such as
nationality or personal diseases history to analyse their impact on
comorbidity.
Another example in this sense is \cite{7}.In \cite{7} the authors analysed the effect of breast cancer treatment
on mortality, taking age at diagnosis and comorbidity into account. The
databases used for this study were 4 nationwide population registries in
Denmark:

\begin{enumerate}
\def\labelenumi{\arabic{enumi}.}
\item
  The Danish Civil Registration System
\item
  The Danish Breast Cancer Cooperative Group
\item
  The Danish National Patient Register
\item
  The Danish Register of Causes of Death provided information on 62 591
  women diagnosed with early-stage breast cancer, in the time range
  1990--2008, about which data on treatment were available for 39 943
  \cite{7}.
\end{enumerate}

To measure comorbidity, the Charlson Comorbidity Index was used, and the
treatments for breast cancer considered were classified as follows:
none, chemotherapy, endocrine therapy and unknown. A multivariable Cox
modelling assessed the effect of comorbidity on breast cancer-specific
mortality, adjusting for age at diagnosis and other clinical factors.
The results show how the impact of comorbidity on mortality was most
present in patients aged 50-79 years, and show how comorbidity at breast
cancer diagnosis is an independent adverse prognostic factor for death
after breast cancer.Also in this case the study is just concentrated in analysing specific
types of comorbidities, and a wider analysis would be interesting.
Moreover it is focused on data coming from a specific nation, and it
could be interesting analysing also data coming from different part of
the world and comparing them to discover other possible relationships.
The method used to extract comorbidities from the data is well explained
by the authors, and validated, therefore the results appear as built on
solid grounds.
Further and more recent studies confirm such trend. 
For instance in \cite{tiruye2024impact} the authors investigates the association between comorbidities and prostate cancer-specific mortality using data from 15,695 South Australian men diagnosed between 2003 and 2019. Comorbidities were measured using the Rx-Risk comorbidity index, and findings indicate that patients with a high comorbidity score (Rx-Risk score greater or equal to 3) had a significantly increased risk of prostate cancer-specific mortality (adjusted sub-hazard ratio [sHR] 1.34). Specifically, medications for cardiac disorders, chronic airway disease, depression and anxiety, and thrombosis were associated with increased mortality risk. Conversely, diabetes and chronic pain medications did not show a significant impact on prostate cancer-specific mortality. The study suggests that managing these comorbid conditions could improve survival outcomes in prostate cancer patients.

Moreover in \cite{forder2023mechanisms} the authors The study analyzed data from men diagnosed with prostate cancer in South Australia between 2003 and 2019. Comorbidity was assessed one year before the prostate cancer diagnosis using the Rx-Risk index. Flexible parametric competing risk regression was employed to estimate the association between comorbidities and prostate cancer-specific mortality, adjusting for sociodemographic variables, tumor characteristics, and treatment types. More specifically Prostate cancer-specific mortality was higher for patients with an Rx-Risk score of 3 or more compared to those with an Rx-Risk score of 0 (adjusted sub-hazard ratio [sHR] 1.34, 95\% CI: 1.15–1.56) and lower comorbidity scores (Rx-Risk scores of 1 or 2) were not significantly associated with prostate cancer-specific mortality.
What can be seen mining the literature is the presence of several studies which relates comorbidities to specific organ cancers, and to specific restricted cohorts of patients \cite{canc1,canc2,canc3}. In such studies it appears clear the variabilities of possible pathological comorbidities appearing in the various cases, with the clear insurgence of a high variability among the several different pathologies. 

\subsection{Comorbidities and Neurological Disorders}
In this section we will propose a set of studies in which comorbidities in neurological related diseases are presented. The set of neurological disorders is vast, and this following list of examples propose itself to be only a starting point for further investigations, and to provide the reader with possible further readings and investigations to be done in these fields.
A first interesting example in this sense is \cite{8}. The main aim of the paper is to describe EpilepsyGene, a database containing
499 genes and 3931 variants associated with 331 clinical epileptic
phenotypes, collected from 818 publications. The data collected were
analysed with in-depth data mining, gaining insights into their
functioning, including functional annotation, gene prioritization,
functional analysis of prioritized genes and overlap focusing on the
comorbidity \cite{8}. Moreover the authors describe the web interface
developed in order to access the various data of interest inside
EpilepsyGene, designed to be a core genetic database to provide
substantial convenience for uncovering the genetic basis of epilepsy.To obtain the list of genes and their functional annotation many tools
and techniques have been applied. Here we will describe the tools and
steps used to obtain information about comorbidity:

\begin{enumerate}
\def\labelenumi{\arabic{enumi}.}
\item
  The initial search for list of genes and mutations relevant to
  epilepsy was done retrospectively through the PubMed database with
  some specific query terms as described in \cite{8}. In particular more
  than 1000 publications, from 1995 to 2014, were obtained. In addition,
  other genetic information relevant to Epilepsy was added from
  pre-existing genetic databases. Genetic data were extracted from the
  articles selected manually, and other information relevant to the
  various data was also collected such as ethnicity, gender (male or
  female), age and inheritance.
\item
  For exploring comorbidities of other disorders with epilepsy, an
  overlap analysis was performed considering the shared genes,
  intersectional epileptic phenotypes and enriched pathways in the genes
  \cite{8}. The relative genes of each disorder were previously collected
  from other databases such as AutismKB \cite{9}, ADHDgene \cite{10} and
  SZGR \cite{11}. These genes collected were compared with the genes in
  EpilepsyGene and as a result 65 genes in EpilepsyGene were found
  shared by MR, 18 by ADHD, 67 by SCZ and 146 by ASD.
\item
  In order to analyse the phenotypes associated with these genes, those
  corresponding to the epileptic genes were retrieved. Moreover pathway
  enrichment analysis for the common genes was undertaken separately.
\end{enumerate}

This paper is interesting since is a comprehensive and collective study
on epilepsy, collecting a large amount of genetic information available
on the topic. The management of epilepsy patients is in fact a critical aspect, is a critical and emerging tools which requires increasing attention from experts and clinical managements in order to provide appropriate support to patients, as described in \cite{peltola2024expert}.
Further several other pathologies have been identified as comorbidities with respect to neurological diseases. For instance fatigue \cite{penner2017fatigue} has been longer debated among being a comorbidity or a concurrent condition. Despite its widespread occurrence across different neurological conditions, the underlying mechanisms of fatigue remain poorly understood. Inconsistencies in how fatigue is defined contribute to its status as one of the least-studied conditions in neurology.
For instance in \cite{roca2019chronic} the authors highlighted how the physiological aging process involves immunosenescence and vascular aging, which can lead to frailty and heightened vulnerability in elderly individuals, and are also implicated in chronic conditions such as stroke, Parkinson's disease, Alzheimer's disease, and peripheral nervous system disorders like polyneuropathy. These neurological disorders play a significant role in the development of geriatric syndromes, which are prevalent among older adults and include multifactorial conditions such as delirium, falls, incontinence, and frailty. Moreover, neurological disorders can both stem from and exacerbate these syndromes, further impacting the overall quality of life, physical function, and health outcomes, including morbidity and mortality. 

\subsection{Text Mining, Machine Learning and Automatic Comorbidities Extraction in Multi-Modal and Complex Diseases}
In most recent years the use of machine learning and deep learning have revolutionized the field of bioinformatics, computer vision and natural language processing \cite{lecun2015deep,dimitri2017drugclust,maj2019integration,spiga2021machine,dimitri2020unsupervised,otter2020survey,khan2021machine}. 
In this context NLP and data mining techniques, as well as the latest advancements in this fields have played a fundamental role in comorbidities mining and understanding. 
For instance in \cite{12} the authors the experiment performed for the workshop and
challenge organized by i2b2 (Informatics for Integrating Biology to the
Bedside). The aim of the challenge was to extract information from
patients discharge summaries about obesity and the comorbidities
associated to it (the various comorbidities could have been: Present,
Absent, Questionable, or Unmentioned in the summary). So the goal of
i2b2b was to provide data for the participants and invite them to
produce automated systems, able to classify obesity and its
comorbidities into four classes based on individual discharge summaries
\cite{12}. The texts proposed to the participants were annotated for textual
judgments, reflecting the explicitly reported information on diseases,
with intuitive judgments reflecting professionals reading of the
information available in the summaries. In total 1237 discharge
summaries were presented to the participants.
The challenge was focused on evaluating the performances of the systems
proposed in this kind of data. Overall there were 30 teams, submitting
two sets of up to three systems runs for evaluation, for a total of 136
submissions.The systems presented were a combination of rule-based and machine
learning approaches \cite{12}.
First of all the texts proposed were annotated, by two obesity experts,
with respect to textual and intuitive annotations (writing which
diseases were explicitly present in the text proposed and which one
could instead be inferred from the data in the text). A third annotator
was asked to solve the conflict in the opinion between the two \cite{12}.
After the annotation phase the data were divided in the training and
test set. The systems proposed were evaluated using micro and macro
averaged precision (P), recall (R) and F measure (F1). Since the
emphasis of the challenge was on the less well represented classes, the
macro-averaged F-measures were used as the main metric for evaluation,
while micro-averaged F-measures were used as a global perspective on the
results \cite{12}.The top 10 systems proposed are summarized and reported in the paper,
and they were ranked according to the measures previously described and
considering the textual and intuitive cases. For example, the best textual system was developed by Yang et Al.
\cite{13}. In their system they used a precompiled dictionary composed by
diseases, symptoms, treatments and medication terms. Having constructed
the dictionary they then looked for sentences matching the terms in the
dictionary. For documents containing more than one sentence about a
disease, they made a weighted combination of the evidence in sentences
determining the class in this way. The second best textual system was developed by Solt et Al. \cite{14}.
They divided discharge summaries into sections, and to mark a disease as
present, they wrote a rule-based classifier with disease names, and
their related terms (such as synonyms or spelling variants). Once
partitioned the text with contextual clues, indicating negative or
uncertain statements, they fed the partitions into a series of binary
classifiers, whose output indicated if the disease was Questionable,
Absent or Present. If no label was assigned to the disease it was signed
as unmentioned. The best intuitive systems, on the other hand, used the output of the
textual systems. In particular many of them determined a default mapping
between the textual and the intuitive systems, and used it as a starting
point for the intuitive judgments. The best four systems developed used
rule-based, incorporating terms highly correlated to the diseases and
symptoms.

This paper presents various interesting results related to the use of
text mining, for extracting biomedical knowledge about comorbidities.
Moreover it shows the difficulty of differentiating textual judgments
from intuitive ones. In particular the overlap in information, currently
used by automated systems for the identification of textual and
intuitive judgments, suggests that the textual judgments of experts in
the domain could differ from the ones of lay people. However even if the
boundaries between textual and intuitive judgments were not clear, it
was interesting to notice how the automated systems performed well in
most pieces of information.

Though we found this study and the methods used really interesting, I
thought the dataset on which the various systems have been tested is
quite small. Therefore, perhaps it could be useful to try to retrieve a
larger dataset and test the various systems on that.
Moreover another interesting paper in this sense is \cite{15}. 
The approach used merges together a variety of automated techniques, and
consists of the following major steps \cite{15}:

\begin{enumerate}
\def\labelenumi{\arabic{enumi}.}
\item
  \textbf{Preprocessing}: this step was implemented with an Automated
  Isolation of Hotspot Passages (AutoHP) and Negation Detection. The
  preprocessing procedure identified and extracted pieces from the text
  that were most likely to bring more information for the classification
  task.
\end{enumerate}

With the AutoHP technique a set of features are taken from a text
collection as input, and they identify those meeting a specific cut-off
value.

\begin{enumerate}
\def\labelenumi{\arabic{enumi}.}
\setcounter{enumi}{1}
\item
  \textbf{Tokenization}: all the passages isolated in the previous step
  by the AutoHP were tokenized into individual features, using the
  algorithm for individual words based on the space and punctuation
  separation. This tokenized set of features was then modeled as a
  binary vector, in which each position corresponded to a feature
  obtained after preprocessing the training set. To each text samples
  there was assigned a binary vector indicating the presence or absence
  of a certain feature.
\item
  \textbf{Vectorization and Filtering}: the vectorization was done with
  a Zero-vector Filtering (ZeroVF). In particular the zero-valued
  feature vectors resulting from the pre-processing were discarded,
  since the assumption was that they contributed for no useful
  information to the classification algorithm. Since from the
  experiments it emerged that the ZeroVF procedure was almost
  universally helpful across comorbidities, it was included in all
  subsequent cross-validation experiments.
\item
  \textbf{Classification}: in both the textual and the intuitive
  problems of classification (see previous section for an explanation of
  their meaning) the classification technique used was ECOC
  (Error-Correcting Output Codes). This approach formulates a multiple
  classification problem, considering a set of several binary
  classification decisions, between all possible subsets of the original
  classes. To classify a new document, each of the sub-classifiers makes
  a prediction and the document is then assigned with respect to the
  class with the most similar result vector. The sub-classification
  problem was obtained using SVM in particular the libSVM library
  \cite{16}.
\end{enumerate}

For each step the best performing procedure was selected using a 5x2 way
cross-validation approach on the training data.

The automated techniques, used for multiple-classification with the
clinical narrative text proposed, demonstrated effectiveness. To have a
more complete idea on the good performances obtained, it could be
interesting to test the techniques on larger datasets. Moreover the
classification task could be performed using different machine learning
approaches and then they could be compared in terms of performances.An interesting and useful point of this implementation is that the
system does not require any a-priori knowledge on the diseases. Furthermore in \cite{17} the authors gathered phenotypic descriptions of
patients form electronic medical, records in order to obtain information
on disease correlations. In particular by extracting phenotype
information, from the free text in the records, it is possible to show
how the information contained in the records can be used to produce a
group division of the patients and disease co-occurrence statistics.The patients records were collected form the the Sct. Hans Mental Health
Centre, in Roskilde, Denmark. In total 5543 were followed within a time
range of 10 years (1998-2008), and all their information stored in an
EPR database.The dictionary employed for the text mining purpose was based on the
Danish translation of the WHO International Classification of Diseases
(ICD10, organized as a hierarchical classification of diseases and
symptoms, and divided into 22 anatomical/functional chapters, increasing
specification of terms going down in the level) downloaded from the
Danish National Board of Health on the 2\textsuperscript{nd} Nov 2009. Given the patient record, the compiled text was normalized for each
patient considering the orthographic variations like in the dictionary.
All the sentences in the records were split in smaller units, and for
each unit an algorithm created all the possible strings of 1-10 words
looking them up in the dictionary and looking for exact matches. The
match chosen was the longest possible one, and polysemic or
misinformative terms were disqualified for classification purposes. To explore comorbidities the authors used two measures. The 226801
possible pairs of different codes were ranked comparing the frequency of
co-occurrence in patients with respect to which would be the correlation
considering no a-priori correlation assumptions. These two measures
ensure statistical significance, when focusing on pairs with an
increased co-association measure. To compute the comorbidity ranking
measures, for each pair of ICD10 codes A and B the patient corpus is
divided in 4 categories: A\&B. A NOT B, B NOT A, and NOT A NOT B
considering their association to A and B. Then using Fisher's exact
test, the pairs are sorted according to their p-value \cite{17}. Moreover, they obtained the comorbidity scores between the different
chapters in ICD10 and considering the comorbidity scores associated to
it. Analysis and methods used in this paper are particularly interesting. In
particular the analysis is extensive and it doesn't focus on a single
disease only, as most of the previous studies presented do. Moreover
authors validate their results in a systematic and complete way.

Further recent studies focused on the use of later machine learning applications for the study of comorbidities and their complex interelationships. 
For instance in \cite{sanchez2019machine} the author addresses challenges in diagnostic associations using machine learning algorithms, highlighting issues like collinearity and variability. To mitigate these, the researchers propose and test the application of uniform manifold approximation and projection (UMAP), a recently popular dimensionality reduction technique. They demonstrate its effectiveness by applying UMAP to a large clinical database of Spanish depression patients, preceding hierarchical agglomerative cluster analysis. Through extensive analysis and validation using unsupervised metrics, they confirm UMAP's ability to reveal consistent patterns that align with established relationships, thereby supporting its utility in advancing the study of comorbidities.
Moreover in \cite{arevalo2022importance} the authors conducted a study which utilized data from the SEMI-COVID-19 Registry in Spain to investigate how different combinations of comorbidities influence outcomes in hospitalized COVID-19 patients. The study applied two machine learning algorithms, Random Forest (RF) and Gaussian mixed model (GMM), to classify comorbidities and patients. The primary endpoint assessed was a composite of all-cause death or intensive care unit admission during hospitalization. The analysis identified several key comorbidities such as heart failure/atrial fibrillation (HF/AF), vascular diseases, and neurodegenerative diseases as most influential in patient outcomes according to the RF algorithm. Clustering analysis revealed six distinct clusters, with clusters 4, 5, and 6 associated with worse outcomes compared to clusters 1, 2, and 3. Cluster 5, characterized by HF/AF, vascular diseases, neurodegenerative diseases, kidney/liver diseases, acid peptic diseases, and chronic obstructive pulmonary disease, showed the poorest prognosis. The study underscores the complex interplay of multiple comorbidities in influencing COVID-19 patient outcomes and complications, highlighting the importance of considering these interactions in clinical management and risk stratification strategies.
Similarly in \cite{aktar2021machine} the authors analysed the impact of comorbities in COVID-19 patients, showing the risk and presence of higher mortality rates,
Morevoer in \cite{uddin2022comorbidity} the authors employed machine learning and network analytics to predict comorbidity and multimorbidity patterns in major chronic diseases. Patient networks were constructed where nodes represented patients and edges indicated shared diseases. Various machine learning models, including Logistic Regression, k-Nearest Neighbors, Naïve Bayes, Random Forest, and Extreme Gradient Boosting, along with deep learning models like Multilayer Perceptrons and Convolutional Neural Networks, were evaluated. Extreme Gradient Boosting achieved the highest accuracy at 95.05\%, followed by Convolutional Neural Networks at 91.67\%. Key findings highlighted the importance of patient trajectory episodes and network transitivity in predicting disease patterns. The study's insights aim to support healthcare stakeholders and policymakers in managing the challenges posed by disease comorbidity and multimorbidity effectively.

\hypertarget{conclusions}{%
\section{Conclusions}\label{conclusions}}
In this brief report and surveyed we started by providing a 4-dimensional definition of the concept of comorbidity. 
Starting from such definition, we presented some examples of
comorbidity mining applications in bio-databases and through text mining
techniques as well as an overview of cancer and neurological diseases comorbidities studies as examples. We showed how the definition of comorbidity is multi-faced and complex, far from being unique and concordant among different studies. With the advent of the most recent Large Language Models and deep learning multimodal applications \cite{dimitri2022short,zhao2023survey,ngiam2011multimodal}, we will probably assist to further applications and knowledge discovery in this field based on the applications of such techniques, which could, potentially, represent a first fundamental and important steps for the progression in this field.

\bibliographystyle{unsrt}  
\bibliography{references}

\end{document}